\documentclass[aps, prl, twocolumn,superscriptaddress,nofootinbib,10pt]{revtex4-2}
\usepackage{graphicx}
\usepackage{dcolumn}
\usepackage{multirow}
\usepackage[table]{xcolor}
\usepackage{colortbl,hhline}
\usepackage{amssymb,amsmath,amsthm,mathrsfs}
\usepackage{epstopdf}
\usepackage{hyperref}
\usepackage{empheq}
\usepackage{color}
\usepackage[normalem]{ulem} 
\usepackage{physics}
\usepackage[normalem]{ulem}
\usepackage[ruled,vlined]{algorithm2e}
\usepackage{enumitem}
\usepackage{empheq}
\usepackage{comment}
\newcommand{\be}{\begin{equation}}
\newcommand{\ee}{\end{equation}}
 \newcommand{\bea}{\begin{eqnarray}}
\newcommand{\eea}{\end{eqnarray}}

\newcommand{\CL}{{\tt ${\mathcal C}$osmo${\mathcal L}$attice}~}

\def \FFdual {F_{\mu\nu}\tilde{F}^{\mu\nu}}

\begin{document}

\title{The strong backreaction regime in axion inflation}

\newcommand{\addressIFIC}{Instituto de F\'isica Corpuscular (IFIC), Consejo Superior de Investigaciones Cient\'ificas (CSIC) and Universitat de Val\`{e}ncia, 46980, Valencia, Spain}
\newcommand{\addressEHU}{Department of Physics, University of Basque Country, UPV/EHU, 48080, Bilbao, Spain}
\newcommand{\addressEHUQC}{EHU Quantum Center, University of Basque Country, UPV/EHU}

\author{Daniel G. Figueroa} \email{daniel.figueroa@ific.uv.es} \affiliation{\addressIFIC} 
\author{Joanes Lizarraga}\email{joanes.lizarraga@ehu.eus} \affiliation{\addressEHU}\affiliation{\addressEHUQC} 
\author{Ander Urio}\email{ander.urio@ehu.eus} \affiliation{\addressEHU}\affiliation{\addressEHUQC} 
\author{Jon Urrestilla}\email{jon.urrestilla@ehu.eus} \affiliation{\addressEHU}\affiliation{\addressEHUQC} 

\date{\today}

\begin{abstract}
We study the non-linear dynamics of axion inflation, capturing for the first time the inhomogeneity and full dynamical range during strong backreaction, till the end of inflation. Accounting for inhomogeneous effects leads to a number of new relevant results, compared to spatially homogeneous studies: {\it i)} the number of extra  efoldings beyond slow roll inflation increases very rapidly with the coupling, {\it ii)} oscillations of the inflaton velocity are attenuated, {\it iii)} the tachyonic gauge field helicity spectrum is smoothed out (i.e.~the spectral oscillatory features disappear), broadened, and shifted to smaller scales, and {\it iv)} the non-tachyonic helicity is excited, reducing the chiral asymmetry, now scale dependent. Our results are expected to impact strongly on the phenomenology and observability of axion inflation, including gravitational wave generation and primordial black hole production. 
\end{abstract}


\maketitle

{\it \underline{Introduction}.--} As inflationary constructions are very sensitive to unknown ultraviolet (UV) physics, a promising candidate for an inflaton is an {\it axion-like particle} that enjoys a shift-symmetry. Possible interactions of such inflaton with other species are then very restricted, protecting in this way the inflationary dynamics from unknown UV physics. While several implementations of axion-driven inflation scenarios have been proposed~\cite{Freese:1990rb, Adams:1992bn,Dimopoulos:2005ac, Easther:2005zr, Bachlechner:2014hsa, McAllister:2008hb, Silverstein:2008sg}, we will simply focus on scenarios where the lowest dimensional shift-symmetric interaction between an inflaton $\phi$ and a hidden Abelian gauge sector, $\phi F_{\mu \nu} \tilde F^{\mu \nu}$, is present, with $F_{\mu \nu}$ the field strength of a {\it dark photon} $A_\mu$, and $\tilde F_{\mu \nu}$ its dual. These scenarios are typically referred to as {\it axion inflation}.

In axion inflation, an exponential production of one of the gauge field helicities is expected during the inflationary period~\cite{Turner:1987vd,Garretson:1992vt,Anber:2006xt, Anber:2009ua,Barnaby:2010vf,Adshead:2013qp,Cheng:2015oqa}. The excited helicity can lead to rich phenomenology such as the production of large density perturbations~\cite{Barnaby:2010vf,Barnaby:2011qe,Barnaby:2011vw,Cook:2011hg, Barnaby:2011qe,Pajer:2013fsa,Domcke:2020zez,Caravano:2022epk} and chiral tensor modes~\cite{Sorbo:2011rz, Barnaby:2011qe, Cook:2013xea,Adshead:2013qp,Bastero-Gil:2022fme,Garcia-Bellido:2023ser}. Such perturbations can be probed by the cosmic microwave background (CMB)~\cite{Barnaby:2010vf,Meerburg:2012id,Sorbo:2011rz}, searches for primordial black holes (PBHs)~\cite{Linde:2012bt, Pajer:2013fsa, Bugaev:2013fya, Cheng:2015oqa, Garcia-Bellido:2016dkw,Garcia-Bellido:2017aan,Domcke:2017fix,Cheng:2018yyr,Ozsoy:2023ryl}, and gravitational wave (GW) detection experiments~\cite{Cook:2011hg,Anber:2012du,Domcke:2016bkh,Bartolo:2016ami}. In addition, fermion production~\cite{Adshead:2015kza,Adshead:2018oaa,Domcke:2018eki}, thermal effects~\cite{Ferreira:2017lnd,Ferreira:2017wlx}, magnetogenesis~\cite{Garretson:1992vt, Anber:2006xt, Adshead:2016iae, Durrer:2023rhc}, baryon asymmetry~\cite{Giovannini:1997eg,Anber:2015yca,Fujita:2016igl,Kamada:2016eeb,Jimenez:2017cdr,Cado:2022evn}, and (p)reheating~\cite{Adshead:2015pva,Adshead:2018doq,Cuissa:2018oiw,Adshead:2019igv,Adshead:2019lbr,Adshead:2019igv,Adshead:2019lbr} mechanisms, can also be efficiently realized. 


{\it \underline{Axion inflation dynamics and methodology}.--}
We consider a total action $S_{\rm tot} = S_{\rm g} + S_{\rm m}$, with standard Hilbert-Einstein {\it gravity} $S_{\rm g} \equiv \int {\rm d}x^4 \sqrt{-g}\,\frac{1}{2}m_p^2 R$, and {\it matter} action ${\cal S}_{\rm m} = -\int {\rm d}x^4 \sqrt{-g}\big\lbrace \frac{1}{2}\partial_\mu \phi\partial^\mu\phi+V(\phi) + \frac{1}{4}F_{\mu\nu}F^{\mu\nu} + \frac{\alpha_{\Lambda}}{4}\frac{\phi}{m_p} \FFdual \big\rbrace$, where $m_p \simeq 2.435\cdot 10^{18}$ GeV is the reduced Planck mass, $\alpha_{\Lambda} \equiv m_p/\Lambda$ the axion-gauge coupling, and $\Lambda$ the axion decay constant. Even though our methodology can be applied to arbitrary potentials, in order to compare with previous results from the literature, we will consider a quadratic potential $V(\phi) = \frac{1}{2}m^2\phi^2$, with $m/m_p \simeq 6.16\cdot10^{-6}$. The variation of $S_{\rm tot}$, specializing the space-time metric to an isotropic and homogeneous spatially flat expanding background, leads to
\begin{empheq}
[left={\empheqlbrace}]{alignat=2}
&\ddot\phi &\,=& -3H\dot\phi+\frac{1}{a^2}\vec\nabla^2\phi-m^2\phi+\frac{\alpha_\Lambda}{a^3 m_p}\vec{E}\cdot\vec{B}\,,\label{eqn:eom1}\\
&\dot{\vec{E}} &=& -H\vec{E}-\frac{1}{a^2}\vec{\nabla}\times\vec{B}-\frac{\alpha_\Lambda}{a m_p}\left(\dot\phi\vec{B}-\vec{\nabla}\phi\times\vec{E}\right),\label{eqn:eom2}\vspace*{2mm}\\
&\ddot{a} &=&
-\frac{a}{3m_p^2}\big( 2\rho_{\rm K}-\rho_{\rm V}+\rho_{\rm EM} \big)\,, 
\label{eqn:eom3}
\end{empheq}
\vspace*{-0.5cm}
\begin{empheq}
[left={\empheqlbrace}]{alignat=2}
\vec{\nabla}\cdot\vec{E} \,\,&&=& -\frac{\alpha_{\Lambda}}{a m_p}\vec{\nabla}\phi\cdot\vec{B}\,,\label{eq:Gauss}\vspace*{2mm}\\
H^2\,&&=& ~\frac{1}{3m_p^2}\big(\rho_{\rm K}+\rho_{\rm G}+\rho_{\rm V}+\rho_{\rm EM}\big)\,,\hspace{1.8cm}
\label{eq:Hubble}
\end{empheq}
with $\dot{} \equiv \partial/\partial t$, $t$ cosmic time, $a(t)$ the scale factor, $H(t)={\dot a/a}$, and where we have defined the magnetic field as $\vec B \equiv \vec\nabla \times \vec A$, the electric field (in the temporal gauge $A_0 = 0$) as $\vec E \equiv \partial_t{\vec A}$, as well as the electromagnetic $\rho_{\rm EM} \equiv \frac{1}{2a^4}\langle a^2\vec E^2+\vec B^2\rangle$ and inflaton's kinetic $\rho_{\rm K} \equiv \frac{1}{2}\langle\dot{\phi}^2\rangle$, potential $\rho_{\rm V} \equiv \langle V \rangle$, and gradient $\rho_{\rm G} \equiv \frac{1}{2a^2}\langle(\vec\nabla\phi)^2\rangle$ homogeneous energy densities, with $\langle ... \rangle$ denoting volume averaging. While~(\ref{eq:Gauss})-(\ref{eq:Hubble}) are constraint equations, Eqs.~(\ref{eqn:eom1})-(\ref{eqn:eom3}) describe the system dynamics, which can be studied under successive levels of approximation:

{\it -- Linear regime}: Deep inside inflation the impact of the gauge field on the inflationary dynamics is negligible, which allows to consistently  neglect the spatial inhomogeneity of the inflaton. However, as the inflaton slowly rolls its potential (we take $\dot\phi < 0$ without loss of generality), the interaction $\dot\phi\vec B$ in Eq.~(\ref{eqn:eom2}) induces an exponential growth in the photon helicity $A_i^{(+)}$, while $A_i^{(-)}$ remains in vacuum. Such chiral instability is controlled by 
\be
\xi=-\frac{\langle\dot\phi\rangle}{2 H \Lambda }\,,\label{eqn:xi}
\ee
so that the gauge field spectrum develops a bump with exponentially growing amplitude, tracking the Hubble scale around $\frac{k}{aH} \sim \frac{1}{\xi}$, for $\xi \gtrsim 1$~\cite{Anber:2009ua}. The linear regime eventually breaks down when the gauge field backreacts on the system, turning  the overall dynamics non-linear. The larger the value of $\alpha_\Lambda$ (or smaller $\Lambda$), the earlier the gauge field backreacts on the dynamics.  

{\it -- Homogeneous backreaction}:  
In this approximation, the backreaction of the gauge field is considered while
enforcing the inflaton to remain homogeneous. This is achieved by neglecting the terms $\propto~\vec\nabla^2\phi$, $\vec{\nabla}\phi\times\vec{E}$, and $\langle(\vec\nabla\phi)^2\rangle$ in Eqs.~(\ref{eqn:eom1}), (\ref{eqn:eom2}) and (\ref{eq:Hubble}), respectively, while promoting, for consistency, $\vec E\cdot \vec B \to \langle \vec E\cdot \vec B \rangle$, in Eq.~(\ref{eqn:eom1}). Even though this regime was originally tackled only approximately, assuming $\dot\phi$ as constant, such limitation was later on surpassed by two methods: {\it i)} solving self-consistently the resulting integro-differential iterative equations~\cite{Cheng:2015oqa,Notari:2016npn,DallAgata:2019yrr,Domcke:2020zez,Peloso:2022ovc}, and {\it ii)} solving the time evolution of the relevant bilinear electromagnetic functions in a {\it gradient expansion} formalism~\cite{Sobol:2019xls,Gorbar:2021rlt}. The two improved methods reached similar conclusions: once 
backreaction becomes relevant, a resonant enhancement of the helical gauge field production is observed, resulting in oscillatory features in the inflaton velocity, as well as in the gauge field spectrum~\cite{Cheng:2015oqa,Notari:2016npn,DallAgata:2019yrr,Domcke:2020zez,Peloso:2022ovc}. This was later understood as due to the time delay between the maximum excitation rate of $A_i^{(+)}$ at slightly sub-Hubble scales, and its backreaction onto the inflaton, dominated by slightly super-Hubble modes~\cite{Domcke:2020zez,Peloso:2022ovc}. 

We remark that in the homogeneous backreaction picture, the gauge field remains {\it maximally helical} ({\it i.e.}~only $A_i^{(+)}$ is exponentially excited), and inflation is sustained for a number of extra efoldings $\Delta \mathcal{N}_{\rm br}$ beyond the would be end of (inflaton driven) slow-roll inflation. 

\begin{figure*}[t]
    \includegraphics[width=1\textwidth]{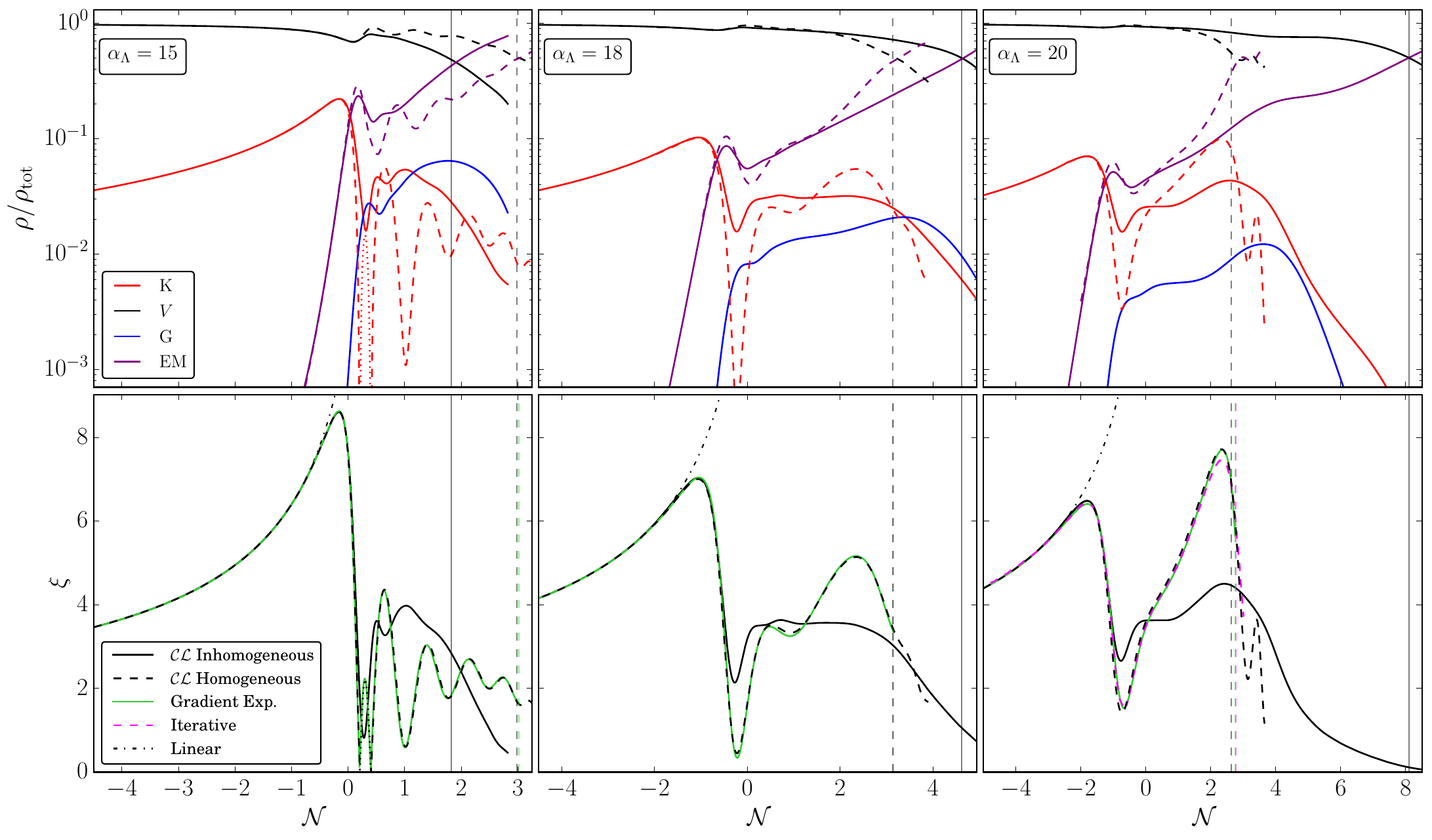}
     \caption{{\it Top Row}: Evolution of the electromagnetic (purple) and inflaton potential (black), kinetic (red) and gradient (blue) energy densities, all normalized to the total energy density of the system, for
     $\alpha_{\Lambda} = 15,\ 18,\ 20$. Solid (dashed) lines correspond to lattice simulations with inhomogeneous (homogeneous) backreaction. 
     {\it Bottom Row}: Evolution of $\xi$  
     for the same coupling constants, corresponding to simulations with inhomogeneous (black solid) and homogeneous (black dashed) backreaction, and to gradient expansion~\cite{Gorbar:2021rlt,Private} (green solid) and iterative method~\cite{Domcke:2020zez} (magenta dashed). Solid and dashed vertical lines signal the end of inflation in each case. Evolution in the linear regime (black dash-dotted) is also shown for completeness.}
     \label{fig:rho_IL15_18}
 \end{figure*}

{\it -- Inhomogeneous backreaction}: In order to address correctly the non-linear dynamics, we need to solve Eqs.~(\ref{eqn:eom1})-(\ref{eqn:eom3}) fully maintaining spatial inhomogeneity, restoring all inflaton gradient terms previously dropped, and using the local expression of $\vec E\cdot\vec B$ for the backreaction. For this, we have implemented in \CL ($\mathcal{CL}$) \cite{Figueroa:2020rrl,Figueroa:2021yhd} a lattice version of Eqs.~(\ref{eqn:eom1})-(\ref{eq:Hubble}), following the lattice gauge-invariant and shift-symmetric formalism of Ref.~\cite{Figueroa:2017qmv,Cuissa:2018oiw}. We use a 2nd order Runge-Kutta time integrator to evolve Eqs.~(\ref{eqn:eom1})-(\ref{eqn:eom3}), monitoring that the constraint Eqs.~(\ref{eq:Gauss})-(\ref{eq:Hubble}) are always verified to better than $\mathcal{O}(10^{-4})$. Details on our lattice formulation can be found in the Supplemental Material and in \cite{AxInflLong}. For an alternative non-shift symmetric lattice formulation, see~\cite{Caravano:2021bfn,Caravano:2022epk}.

We start our simulations in the linear regime, with all comoving modes captured  between the infrared (IR) and UV lattice cutoff scales, $k_{\rm IR}  \leq k \leq k_{\rm UV}$, well inside the initial comoving Hubble radius $1/aH$. By setting initially $k_{\rm IR}/(a H) \simeq 10$, all gauge field modes of both helicities are initialized in a {\it Bunch-Davies} (BD) quantum vacuum state $A^{(\pm)} \simeq e^{i k/aH}/\sqrt{2k}$. The initial fluctuations serve then as  a seed for the tachyonic instability of $A_i^{(+)}$: as the modes approach the Hubble scale, their amplitude starts growing exponentially. In order to capture the dynamics correctly, we first solve, in the lattice, the linear regime of the gauge field, up to a given cut-off $k < k_{\rm BD}$, with $k_{\rm IR} \ll k_{\rm BD} \ll k_{\rm UV}$. We let the most IR modes grow till they dominate over the BD tail within the range $k_{\rm IR} \leq k < k_{\rm BD}$. Then, we {\it switch} to evolve the non-linear Eqs.~(\ref{eqn:eom1})-(\ref{eqn:eom3}), allowing all fields to be excited in the full lattice range $k \in [k_{\rm IR}, k_{\rm UV}]$. After the switch, the system still remains in the linear regime for a while (coupling dependent), until  the backreaction of the gauge field becomes noticeable on both the inflaton and the expansion dynamics. From that moment the system dynamics becomes fully non-linear, entering into the strong backreaction regime for sufficiently large couplings. 

\begin{figure}[t]
\hspace*{-0.25cm}\includegraphics[width=8.8cm]{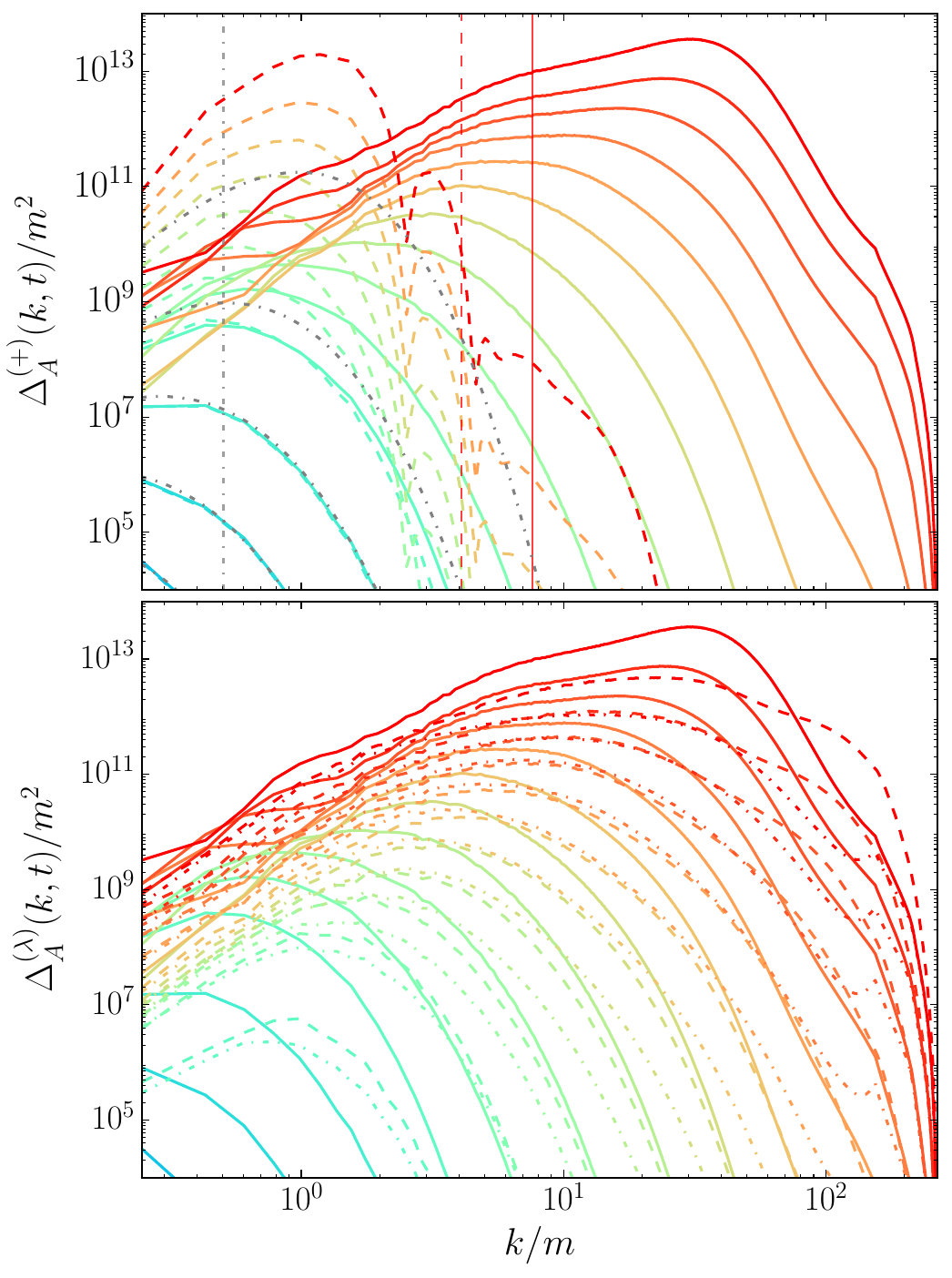}
\caption{\label{fig:PS_IL18}Evolution the gauge field power spectra for $\alpha_{\Lambda} = 18$. \textit{Top:} $\Delta_A^{(+)}(k,t)$ spectra from simulations in the linear regime (gray dash-dotted lines), and with homogeneous (dashed lines) and inhomogeneous (solid lines) backreaction. Vertical lines represent the comoving Hubble scale at the end of inflation in each case. \textit{Bottom:} Different gauge polarization power spectra from a simulation with inhomogeneous backreaction: $\Delta_A^{(+)}(k,t)$ (solid lines), $\Delta_A^{(-)}(k,t)$ (dash-dotted lines) and $\Delta_A^{(\rm L)}(k,t)$ (dashed lines). In all panels, lines are separated by $\Delta \mathcal{N} = 0.5$ from earlier times to later ones, from colder to hotter, except in the linear regime. The reddest spectrum corresponds to the end of inflation for each case.}
\end{figure}

{\it \underline{Results}.--}
In the following, we present our study on the strong backreaction regime, which requires $\alpha_{\Lambda}\gtrsim 15$, 
capturing the inhomogeneity and full dynamical range of the system, until the end of inflation. A detailed description of our procedure and results will be presented in~\cite{AxInflLong}.

We list our run parameters in Table~\ref{tab:sims}, where $N$ is the number of lattice sites per dimension, $\tilde{L}=mL$ the comoving lattice length, $\kappa_{\rm UV}=k_{\rm UV}/m$ the lattice UV scale, $\kappa_{\rm BD}$ the BD cut-off scale (set by trial and error), $\mathcal{N}_{\rm start}$ the number of efolds before the end of slow-roll inflation (marked as $\mathcal{N}=0$) when we start our simulation, and $\mathcal{N}_{\rm switch}$ the moment when all inhomogeneous terms are activated. For convenience we set $a=1$ at $\mathcal{N}=0$.

\begin{table}[h!]
\renewcommand{\arraystretch}{1.35}
\resizebox{0.9\columnwidth}{!}{\begin{tabular}{|c||c|c|c|c|c|c|}
\hline
  & $\boldsymbol{N}$ & $\boldsymbol{\tilde{L}}$ & $\boldsymbol{\kappa_{\rm UV}}$ & $\boldsymbol{\kappa_{\rm BD}}$ & $\boldsymbol{\mathcal{N}_{\rm start}}$ & $\boldsymbol{\mathcal{N}_{\rm switch}}$\\ \hline
 {} & {} & {} & {}& {} & {} & {}\\[-3.25ex]\hline \\[-3.5ex]
  \,$\boldsymbol{\alpha_{\Lambda}=15}$\, & \,640\, & \,32.524\, & \,106.981\, & 46 & -4.5 & -1.1 \\ \cline{1-7}
  \,$\boldsymbol{\alpha_{\Lambda}=18}$\, & \,1600\, & \,32.524\, & \,267.594\, & 10 & -4.5 & -1.8 \\ \cline{1-7}
  \,$\boldsymbol{\alpha_{\Lambda}=20}$\, & \,2340\, & \,50.971\, & \,170.746\, & 9 & -5 & -2.4 \\ \hline
\end{tabular}}
 \caption{\label{tab:sims} Parameters used in the simulations.}
\end{table}

\begin{figure}[t]
\includegraphics[width=8.8cm]{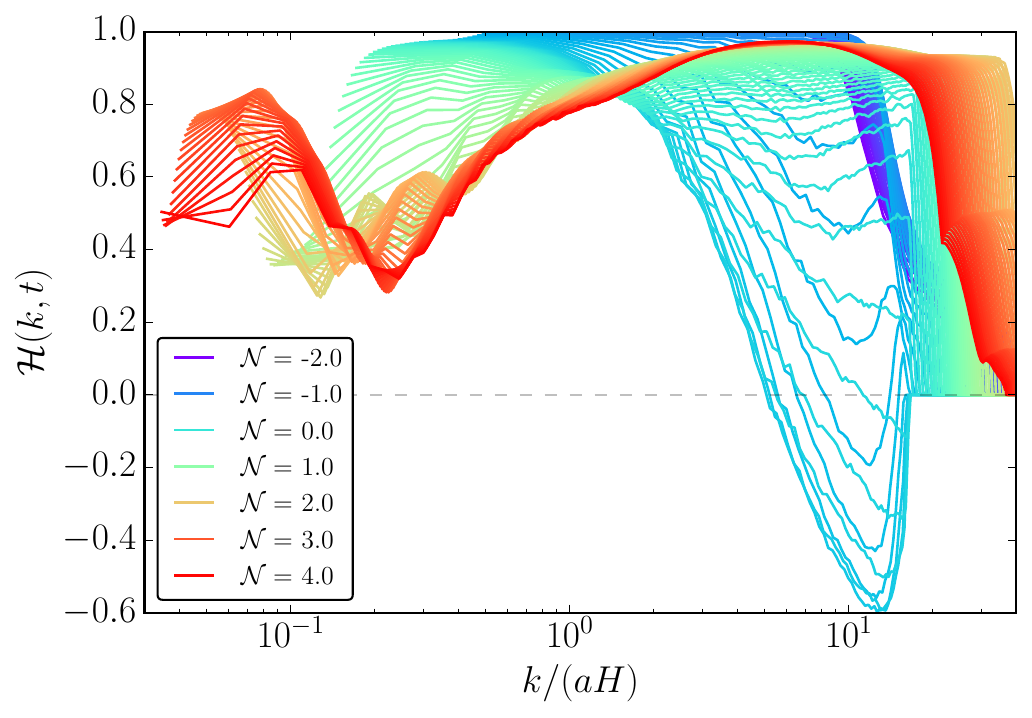}
\caption{Non-linear evolution of the {\it normalized spectral helicity} as defined in Eq.~(\ref{eq:SpecHel}) vs. $k/(aH)$ for $\alpha_{\Lambda}=18$. Colour coding goes from earliest (colder) to latest (hotter) times in the simulation. We start plotting from $\mathcal{N}_{\rm switch}$ onwards and the separation between different lines is $\Delta \mathcal{N} = 0.05$ efoldings.}
\label{fig:SpectralHelicity}
 \end{figure}

Our results are summarized by a series of figures, where we compare the outcome of our simulations for the linear,  homogeneous  backreaction, and  inhomogeneous  backreaction regimes. In the top panel of Fig.~\ref{fig:rho_IL15_18}, we plot the evolution of the electromagnetic and inflaton's kinetic, gradient and potential homogeneous energy densities (normalized by the total energy density), whereas in the bottom panel, we show the evolution of $\xi$, c.f.~Eq.~(\ref{eqn:xi}). In both panels we show, for each coupling  considered $\alpha_{\Lambda}$, the system evolution as a function of the number of efoldings $\mathcal{N}$, from the initial moment of the simulation in the linear regime, till the end of inflation in the strong backreaction regime. While $\mathcal{N} = 0$ signals the end of slow-roll inflation, the dashed and solid vertical lines indicate the end of inflation, identified as $\epsilon_H \equiv -\dot H/ H^2 = 1$, according to the homogeneous and inhomogeneous backreaction regimes, respectively. Whenever possible, we compare with the outcome from the gradient expansion formalism~\cite{Gorbar:2021rlt,Private} and from the iterative method~\cite{Domcke:2020zez}. Incidentally, our code reproduces accurately the linear and homogeneous backreaction regimes in their corresponding limits, confirming the validity of the code.

We define the power spectrum of the gauge field as $\Delta_{A}^{(\lambda)}(k,t) \equiv \frac{k^3}{2\pi^2}\mathcal{P}^{(\lambda)}_{A}(k,t)$, where $\langle \vec{A}^{(\lambda)}({\vec{k}},t)\vec{A}^{(\lambda')*}({\vec{ k}}',t) \rangle$ $\equiv (2\pi)^3 \mathcal{P}_{A}^{(\lambda)}(k,t)\delta_{\lambda\lambda'}\delta_{\rm D}({\vec {k}}-{\vec{k}}')$ represents an ensemble average. In Fig.~\ref{fig:PS_IL18} we plot various power spectra for a fiducial value $\alpha_\Lambda = 18$, and compare the outcome of our inhomogeneous treatment against the solutions of the homogeneous backreaction and linear regimes. In Fig.~\ref{fig:SpectralHelicity} we also show the helicity imbalance measured through a {\it normalized spectral helicity} observable defined as 
\begin{eqnarray}
\mathcal{H}(k,t) \equiv \frac{\Delta_A^{(+)}-\Delta_A^{(-)}}{\Delta_A^{(+)}+\Delta_A^{(-)}}\,.
\label{eq:SpecHel}
\end{eqnarray}

The inclusion of the inhomogeneous terms brings considerable novelties into the dynamics:

{\it 1.-} The gauge energy $\rho_{\rm EM}$ grows exponentially fast during the linear regime, until it reaches a ${\it few}~\%$ of $\rho_{\rm K}$. The latter, that had been previously slowly growing on a slow-roll trajectory, starts then decreasing, signaling the onset of backreaction. In the homogeneous case, $\rho_{\rm EM}$ and $\rho_{\rm K}$  may perform some large oscillations~\cite{Domcke:2020zez,Peloso:2022ovc}, almost in opposite phase. Such oscillations are however damped in the inhomogeneous dynamics, where the gradient energy $\rho_{\rm G}$ is also significantly excited, with its contribution potentially comparable or even higher than $\rho_{\rm K}$. This could never be captured in the homogeneous regime, where by construction $\rho_{\rm G} = 0$. In the homogeneous case, for some couplings ({\it e.g.} $\alpha_{\Lambda} = 15$) the first and largest oscillation leads $\langle\dot\phi\rangle$ to even flip its sign, with $\xi$ crossing zero back and forth (depicted in the figure by dotted lines), signaling that the inflaton climbs its own potential. This, however, never happens in the inhomogeneous case, where the growth of $\rho_{\rm G}$ damps the oscillation amplitude, and prevents $\xi$ from becoming negative.

{\it 2.-} For all couplings considered, either in the homogeneous or inhomogeneous regimes, inflation ends when $\rho_{\rm EM}$ becomes comparable to $\rho_{\rm V}$, resulting in a Universe already reheated at that moment, which is actually consistent with previous preheating studies for $\alpha_\Lambda \lesssim 15$~\cite{Adshead:2015pva,Adshead:2018doq,Cuissa:2018oiw,Adshead:2019igv,Adshead:2019lbr}. In the homogeneous case, the number of extra efoldings is $\Delta \mathcal{N}_{\rm br} \approx 3$ for all couplings considered. In contrast, in the inhomogeneous dynamics, the number of extra efoldings grows strongly and monotonically with $\alpha_{\Lambda}$, from $\Delta \mathcal{N}_{\rm br} \approx 2$ for $\alpha_\Lambda = 15$ to $\Delta \mathcal{N}_{\rm br} \approx 8$ for $\alpha_\Lambda = 20$.  For $\alpha_{\Lambda}=15$ inflation actually ends earlier in the inhomogeneous case than in the homogeneous regime, whilst for $\alpha_{\Lambda}=18$ and $\alpha_{\Lambda}=20$ it ends later and much later, respectively. The larger the coupling, the earlier backreaction happens ($\rho_{\rm EM}$ surpassing $\rho_{\rm K}$), as expected, roughly at the same time in both approaches. In the inhomogeneous case, the earlier the crossover happens, the longer inflation is prolonged in a quasi-de Sitter regime dominated by $\rho_V$ and $\rho_{\rm EM}$.

{\it 3.-} In the linear regime, the power spectrum of the unstable helicity $\Delta_{A}^{(+)}(k,t)$ develops an exponentially growing peak, tracking the Hubble scale at $k/a \sim H/\xi$. However, as the top panel of Fig.~\ref{fig:PS_IL18} shows, the shape of the power spectra changes considerably when backreaction is considered. In the homogenoeus case (dashed), the spectrum peak grows resonantly in amplitude once backreaction starts, but shifts mildly its (slightly) super-horizon position reached at the onset of backreaction. The spectrum also develops an oscillatory pattern at scales around the Hubble radius in its UV tail. In the inhomogeneous case (solid), on the contrary, oscillatory features are never imprinted in the spectrum, which now spreads power into UV scales, shifting gradually its peak to smaller (slightly sub-horizon) scales, as inflation carries on. As a result the spectrum becomes smoother and wider. The homogeneous and inhomogeneous spectra actually demonstrate that the two approaches capture very different physics.

{\it 4.-} The bottom panel of Fig.~\ref{fig:PS_IL18} features yet another new result due to including inhomogeneity. As the inflaton gradients are developed, the terms $\propto \vec{\nabla}\phi\times\vec{E}$ in Eq.~(\ref{eqn:eom2}) drive the excitation of the \textit{longitudinal} mode $A_i^{\rm (L)}$, as well as of the other circular polarization $A_i^{(-)}$, which had previously remained in vacuum. Furthermore, the term $\propto \dot{\phi}\vec{B}$ also contributes to stimulate $A_i^{(-)}$, thanks to the inhomogeneity of $\dot\phi$. When we switch our simulations to an evolution with Eqs.~(\ref{eqn:eom1})-(\ref{eqn:eom3}), $A_i^{\rm (L)}$ and $A_i^{(-)}$ start with a non-vanishing amplitude much smaller than $A_i^{(+)}$. However, towards the end of inflation, once strong backreaction is at play, $A_i^{(-)}$ and $A_i^{\rm (L)}$ become comparable to (when not even larger than) $A_i^{(+)}$, depending on the scale. To quantify this result, we plot in Fig.~\ref{fig:SpectralHelicity} the spectral helicity [c.f.~Eq.~(\ref{eq:SpecHel})] for our fiducial choice $\alpha_\Lambda = 18$. Whereas in the homogeneous case the gauge field excitation is maximally chiral ($\mathcal{H}(k,t) = 1$), this is no longer the case when all fields are allowed to fluctuate locally. For instance, Fig.~\ref{fig:SpectralHelicity} shows that at the end of inflation, $\Delta_A^{(+)} \approx 3 \Delta_A^{(-)}$ ({\it i.e.}~$\mathcal{H}(k,t) \approx 1/2$) at slightly super-Hubble scales, whereas the helicity balance tends to be restored ($\mathcal{H}(k,t) \rightarrow 0$) in the UV tail of the spectrum, at around $k/a \gtrsim 10H $. Remarkably, the evolution around $\mathcal{N}\sim 0$ shows that $A_i^{(-)}$ dominates over $A_i^{(+)}$ at ${k/a} \sim 10H$, with $\mathcal{H}(k,t) \gtrsim -1/2$. We shall discuss further the excitation mechanism of $A_i^{\rm (L)}$ and $A_i^{(-)}$ in~\cite{AxInflLong}. We note that analogous helicity restoration effects at sub-horizon scales have also been reported in preheating studies~\cite{Adshead:2015pva,Adshead:2016iae}, for the milder coupling regime $9 \lesssim \alpha_\Lambda \lesssim 14$.

{\it \underline{Discussion}}.-- Observable CMB scales leave the Hubble radius during inflation, when the gauge dynamics is well described by the linear regime, and backreation is negligible. Backreaction becomes typically important towards the end of inflation, when large tensor~\cite{Sorbo:2011rz, Barnaby:2011qe, Cook:2013xea,Adshead:2013qp,Anber:2012du,Domcke:2016bkh,Bartolo:2016ami,Bastero-Gil:2022fme,Garcia-Bellido:2023ser} and scalar~\cite{Barnaby:2010vf,Barnaby:2011qe,Barnaby:2011vw,Cook:2011hg, Barnaby:2011qe,Linde:2012bt,Pajer:2013fsa,Bugaev:2013fya,Garcia-Bellido:2016dkw,Garcia-Bellido:2017aan,Domcke:2017fix,Cheng:2018yyr,Caravano:2022epk,Ozsoy:2023ryl} perturbations can be generated. These can lead to  potentially observable quantities, such as a population of PBHs and a stochastic background of GWs, both crucial predictions to probe axion inflation scenarios. Therefore, it is of the utmost importance to describe correctly the system dynamics when backreaction cannot be neglected. 

In this Letter we report the results of using a gauge-invariant and shift-symmetric lattice formalism, capturing for the first time the inhomogeneity and full dynamical range during strong backreaction, till the end of inflation. We explore the parameter space $\alpha_{\Lambda} \gtrsim 15$, which has never been studied during the whole inflationary period while incorporating inhomogenous effects. Such large coupling regime is crucial to understand the generation of scalar perturbations during inflation, which later on lead to PBH formation. While GW production during preheating constrains the coupling down to $\alpha_{\Lambda}\lesssim 15$~\cite{Adshead:2019igv,Adshead:2019lbr}, this depends on the details of the last stages of inflation and of a potential early PBH dominated phase ensued after inflation~\cite{Domcke:2020zez}. As the strong backreaction inflationary phenomenology uncovered in our work is (likely) expected to affect this limit, the exploration of couplings beyond current preheating bounds becomes well justified and crucial to understand observational constraints of axion inflation.

One of the most relevant aspects of our results is the observed `{\it exponential UV sensitivity}' of the dynamics to small coupling increments. In particular, as longer inflationary periods emerge for larger couplings, successively smaller scales need to be resolved. Our simulation data show that when UV scales are not properly resolved, neither the width nor the peak location of the gauge spectra are well obtained (a detailed IR/UV lattice study to highlight this aspect will be presented in~\cite{AxInflLong}). As a simultaneous capture of IR and UV scales is required, this is ultimately the reason why we limited our current study to $\alpha_{\Lambda} \leq 20$, as $\alpha_{\Lambda} = 20$ already required $N > 2300$ sites/dimension to capture correctly all IR/UV scales. Our results show that a correct description of the dynamics can only be provided if inhomogeneities are completely resolved at all scales of interest. To explore larger couplings, much larger lattices will then be needed. In this respect, we notice that the study of the strong backreaction regime for $\alpha_{\Lambda} = 25$ by Ref.~\cite{Caravano:2022epk}, given the lattice sizes reported, cannot capture the full dynamical range required to characterize the non-linear dynamics till the end of inflation.

To summarize our findings we note that the effect of the inhomogeneity is highly non-trivial and requires a dedicated study for each coupling. In general, the excitation and backreaction of the gauge field is no longer controlled by a homogeneous $\xi$ parameter, and resonant oscillatory backreaction features reported by previous homogeneous analyses~\cite{Domcke:2020zez,Gorbar:2021rlt,Peloso:2022ovc}, are quite attenuated. The resulting gauge field spectra during inhomogeneous backreaction become smoother than in the homogeneous case, as no spectral oscillatory features are developed. Furthermore, gauge spectra become wider, spreading power into shorter scales, as the peak spectrum trails the Hubble scale during the $\Delta{\mathcal N}_{\rm br}$ extra efoldings, which grows very strongly with the coupling.

We conclude that the novelties of consistently taking into account the inhomogeneity of the system during strong backreaction will inevitably have an impact on the properties of the scalar and tensor perturbations derived considering homogeneous backreaction, e.g.~\cite{Domcke:2020zez,Bastero-Gil:2022fme,Garcia-Bellido:2023ser}. Furthermore, the completely new feature of scale-dependent gauge chirality makes the possibility of probing these scenarios through their observational windows even more interesting. The observability and phenomenology of axion inflation scenarios will require a complete revision of the state-of-the-art predictions, which we plan to address in future work.

\acknowledgments

\textit{Acknowledgements.-} We thank V.~Domcke, Y.~Ema, S.~Sandner, K.~Schmitz and O.~Sobol for discussion and for kindly providing output data from the gradient expansion formalism. We are equally grateful to R.~Durrer for discussion and comments, and to Z.~Weiner for constructive criticism. We also thank J.~Baeza-Ballesteros for helping us with the launch of simulations in the Lluis Vives cluster. DGF (ORCID 0000-0002-4005-8915) is supported by a Ram\'on y Cajal contract with Ref.~RYC-2017-23493. This work was supported by Generalitat Valenciana grant PROMETEO/2021/083, and by  Spanish Ministerio de Ciencia e Innovaci\'on grant PID2020-113644GB-I00. JL (ORCID 0000-0002-1198-3191), AU (ORCID 0000-0002-0238-8390) and JU (ORCID 0000-0002-4221-2859) acknowledge support from Eusko Jaurlaritza (IT1628-22) and by the PID2021-123703NB-C21 grant 
funded by MCIN/AEI/10.13039/501100011033/ and by ERDF; ``A way of making Europe”. In particular, AU gratefully acknowledges the support from the University of the Basque Country grant (PIF20/151).  This work has been possible thanks to the computing infrastructure of the ARINA cluster at the University of the Basque Country, UPV/EHU, and Lluis Vives cluster at the University of Valencia.

\bibliography{automatic,manual}

\widetext
\newpage
\section{Supplemental Material: lattice discretisation}
\setcounter{figure}{0}
\renewcommand{\thefigure}{S\arabic{figure}}

The lattice discretisation of the axion inflation model has been done following the prescription of \cite{Figueroa:2017qmv,Cuissa:2018oiw} for the spatial discretisation, which is a formalism that preserves gauge-invariance and shift-symmetry exactly on the lattice. We assume that the scalar field $\phi$ lives at lattice sites $\textbf{n}$, whereas gauge fields $A_i$ live at the links between lattice sites, at $\textbf{n}+\hat{\imath}/2$. The spatial and temporal derivatives are the usual forward/backward lattice derivatives: $\Delta^{\pm}_{\mu} \varphi \equiv \frac{\pm 1}{dx^{\mu}}(\varphi_{\pm\hat{\mu}}-\varphi)$, with $dx^{\mu}$ a derivative step, and $\pm \hat{\mu}$ subscripts a unitary displacements in the direction $\hat{\mu}$.

We have used the following lattice definitions of the electric and magnetic fields
\be
E_i(\textbf{n}+\hat{\imath}/2) \equiv \Delta_0^+A_i\, , \quad B_i(\textbf{n}+\hat{\imath}/2+\hat{\jmath}/2) \equiv \sum_{j,k}\epsilon_{ijk}\Delta^+_jA_k\, ,
\ee
and \textit{improved} versions,
\be
 E_{i}^{(2)} (\textbf{n}) \equiv \frac{1}{2}\left(E_{i}+E_{i,-\hat{\imath}}\right)\; , \quad B_{i}^{(4)} (\textbf{n})\equiv \frac{1}{4}\left(B_{i}+B_{i,-\hat{\jmath}}+B_{i,-\hat{k}}+B_{i,-\hat{\jmath}-\hat{k}}\right)\; ,
\ee
for which we explicitly indicate where they live in the lattice.

We use the number of efoldings of the scale factor as the natural time variable. The change of variables from the cosmic time reads
\be
d\mathcal{N} = H dt\; ,
\ee
promoting the Hubble rate $H$ as a dynamical variable, while the scale factor is given by $a=a_ie^{\mathcal{N}-\mathcal{N}_i}$, with $a_i$ the scale factor at some reference time $\mathcal{N}_i$. We choose $a_i=1$ at $\mathcal{N}_i=0$.

We operate in the following set of dimensionless spacetime and field variables called \textit{program variables}, which are defined in terms of the axion mass $m$ as
\begin{equation}
   d\tilde{x}^{\mu}=m dx^{\mu}\;, \quad \tilde{\phi}=\frac{\phi}{m} \;, \quad \tilde{A}_{\mu}=\frac{A_{\mu}}{m}\;, \quad \tilde{H} = \frac{H}{m}\;.
\end{equation}

The lattice version of the equations of motion (\ref{eqn:eom1})-({\ref{eqn:eom3}}) can then be written as
\begin{eqnarray}
&&\tilde{\phi}''=-3\tilde{\phi}' + \frac{1}{\tilde{H}}\left(\frac{1}{a^2}\sum_{i} \tilde{\Delta}_{i}^{-}\tilde{\Delta}_{i}^{+}\tilde{\phi} - \tilde{\phi}+ \frac{\alpha_{\Lambda}}{2a^{3}}\frac{m}{m_{p}}\sum_{i}\tilde{E}_{i}^{(2)}\tilde{B}_{i}^{(4)}\right)\, , \\
&&\tilde{E}'_{i}=-\tilde{E}_{i} + \frac{1}{\tilde{H}}\left( -\frac{1}{a^2}\sum_{j,k}\epsilon_{ijk}\tilde{\Delta}_{j}^{-}\tilde{B}_{k} - \frac{\alpha_{\Lambda}}{2a}\frac{m}{m_{p}}\left(\tilde{\phi}'\tilde{B}_{i}^{(4)} + \tilde{\phi}'_{+\hat{\imath}}\tilde{B}_{i,+\hat{\imath}}^{(4)}\right)\right. \nonumber \\
&&\left.+\frac{\alpha_{\Lambda}}{4a }\frac{m}{m_{p}}\sum_{\pm}\sum_{j,k}\epsilon_{ijk}\left\lbrace \left[(\tilde{\Delta}_{j}^{\pm}\tilde{\phi})\tilde{E}^{(2)}_{k,\pm \hat{\jmath}}\right]_{+\hat{\imath}}+\left[(\tilde{\Delta}_{j}^{\pm}\tilde{\phi})\tilde{E}^{(2)}_{k,\pm \hat{\jmath}}\right]\right\rbrace\right)\, , \\
&&\tilde{H}'= -\frac{1}{3m^2_{p}\tilde{H}}\left(3\tilde{\rho}^{L}_{\rm K}+\tilde{\rho}^{L}_{\rm G}+2\tilde{\rho}^{L}_{\rm EM}\right)\, ,
\end{eqnarray}
while the constraint Eqs.~(\ref{eq:Gauss})-({\ref{eq:Hubble}) read
\begin{eqnarray}
 &&\sum_{i}\tilde{\Delta}_{i}^{-}\tilde{E}_{i} = -\frac{\alpha_{\Lambda}}{2a}\frac{m}{m_{p}}\sum_{\pm}\sum_{i}\left(\tilde{\Delta}_{i}^{\pm}\tilde{\phi} \right)\tilde{B}_{i,\pm \hat{\imath}}^{(4)} \; , \label{eq:GaussLat} \\
 &&\tilde{H}^2=\frac{1}{3 m^2_{p}}(\tilde{\rho}^{L}_{\rm K}+\tilde{\rho}^{L}_{\rm G}+\tilde{V}^{L}+\tilde{\rho}^{L}_{\rm EM})\; , \label{eq:HubbleLat} 
\end{eqnarray}
where we have used that $' \equiv d/d\mathcal{N}$. The lattice version of the homogeneous energy density components is
\begin{equation}
    \tilde{\rho}^{L}_{\rm K}=\frac{\tilde{H}}{2}\left\langle\tilde{\phi}'^{2}\right\rangle_{L}\;, \quad  \tilde{\rho}^{L}_{\rm G}=\frac{1}{2a^2}\left\langle \sum_{i} (\tilde{\Delta}^{+}_{i} \tilde{\phi})^2 \right\rangle_{L}\;, \quad \tilde{V}^{L}=\frac{1}{2}\left\langle \tilde{\phi}^2 \right\rangle_{L}\;, \rho_{\rm EM} = \frac{1}{2a^4}\left\langle \sum_{i}(a^2E^{2}_{i} + B^{2}_{i}) \right\rangle_{L}\;,
\end{equation}
with $\langle ... \rangle_{L} \equiv \frac{1}{N^3} \sum_{\textbf{n}}(...)$ representing lattice volume averaging.

In order to check the level at which the lattice constraints of Eqs.(\ref{eq:GaussLat})-(\ref{eq:HubbleLat}) are obeyed, we propose a couple of dimensionless observables. For the energy conservation of Eq.~(\ref{eq:HubbleLat}) we use the following definition,

\begin{equation}
    \Delta_H = \frac{|LHS-RHS|}{\sqrt{LHS^2+RHS^2}}\; , \label{eq:HubbleSch} 
\end{equation}
whereas for the Gauss constraint Eq.~(\ref{eq:GaussLat}), we use
\begin{equation}
     \Delta_G = \frac{\left\langle|LHS-RHS|\right\rangle_{L}}{\left\langle\sqrt{(LHS_1)^2+(LHS_2)^2+(LHS_3)^2+RHS^2}\right\rangle_{L}}\; .
     \label{eq:GaussSch} 
\end{equation}
In both cases $LHS$ and $RHS$ refer to the {\it left-} and {\it right-hand sides} of the corresponding equation, and $LHS_i=\tilde{\Delta}_{i}^{-}\tilde{E}_{i}$ (considering no sum over repeated indices).

In Fig.~\ref{fig:Constraints} we show the evolution of both constraints for $\alpha_{\Lambda} = 15$, $18$ and $20$.
 \begin{center}
\begin{figure}[h]
    \centering
    \includegraphics[width=1\textwidth]{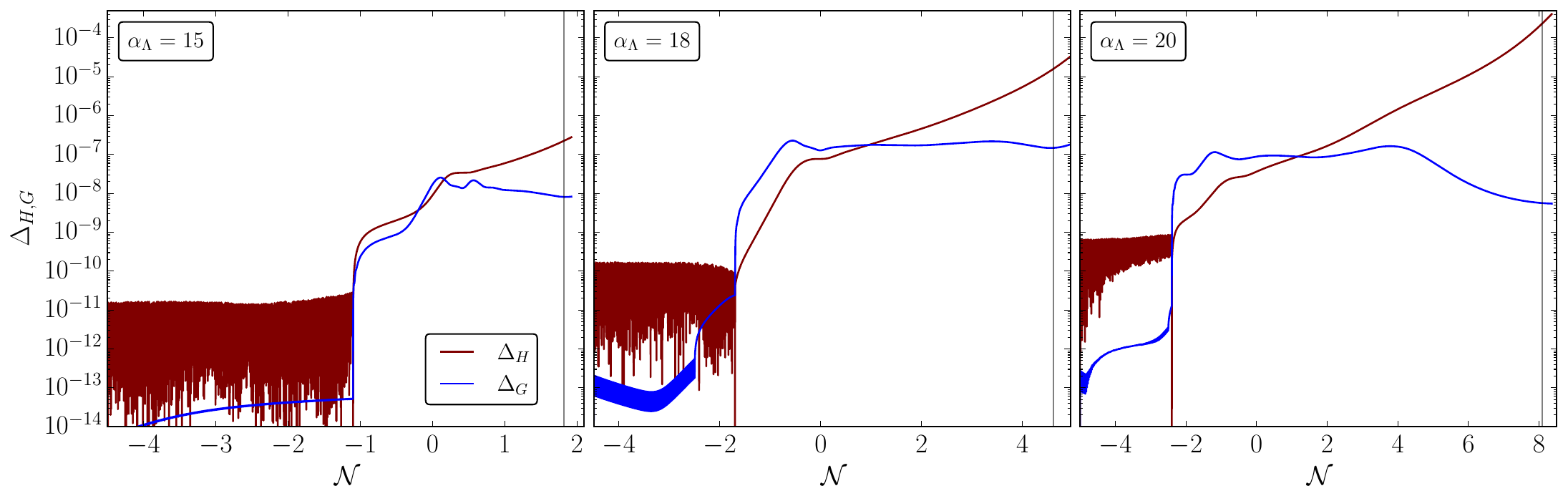}
     \caption{\label{fig:Constraints} Energy conservation (brown lines) and Gauss constraint conservation (blue lines) levels as measured by Eqs.~(\ref{eq:HubbleSch}) and (\ref{eq:GaussSch}) for  $\alpha_{\Lambda} = 15$ (left), $18$ (central) and $20$ (right).}
 \end{figure}
\end{center}

\end{document}